\documentclass[twocolumn]{aastex631}
\usepackage{float}
\usepackage{multirow}
\usepackage[para]{threeparttable}
\usepackage{enumitem}
\usepackage{hyperref}
\usepackage{graphicx}
\usepackage{rotating}
\let\tablenum\relax
\usepackage{siunitx}

\begin{document}

\title{On the Border: Searching for Cometary Activity Near the Centaur-JFC Transition Line}

\author[0000-0003-4094-9408]{A. Fraser Gillan}
\affiliation{Astrophysics Research Centre, School of Mathematics and Physics, Queen’s University Belfast, Belfast, BT7 1NN, UK}
\affiliation{Astrophysics Division, National Centre for Nuclear Research, Pasteura 7, 02-093 Warsaw, Poland}

\author[0000-0003-0250-9911]{Alan Fitzsimmons}
\affiliation{Astrophysics Research Centre, School of Mathematics and Physics, Queen’s University Belfast, Belfast, BT7 1NN, UK}

\author[0000-0001-7335-1715]{Colin Orion Chandler}
\affiliation{Dept. of Astronomy \& the DiRAC Institute, University of Washington, 3910 15th Ave NE, Seattle, WA 98195, USA}
\affiliation{LSST Interdisciplinary Network for Collaboration and Computing, 933 N. Cherry Avenue, Tucson, AZ 85721, USA}
\affiliation{Dept. of Astronomy \& Planetary Science, Northern Arizona University, PO Box 6010, Flagstaff, AZ 86011, USA}

\author[0000-0001-9328-2905]{Colin Snodgrass}
\affiliation{Institute for Astronomy, University of Edinburgh, Royal Observatory, Edinburgh, EH9 3HJ, UK}

\author[0000-0001-9505-1131]{Joseph Murtagh}
\affiliation{Astrophysics Research Centre, School of Mathematics and Physics, Queen’s University Belfast, Belfast, BT7 1NN, UK}

\begin{abstract}    
Current wide-field surveys discover $\sim$15 Jupiter-family comets (JFCs) each year, typically identified via visual detection of a dust coma or tail. The same surveys also discover many asteroids that have distant JFC-like orbits, but with no reported activity. We observed asteroids on Jupiter-crossing orbits beyond the depth of typical survey imaging using the 2.5 m Isaac Newton Telescope. We used deep imaging to observe 16 asteroids in this region, plus 7 known comets for comparison. Three asteroids (2011 WM183, (669525) 2012 XO144, and 2020 RX133) showed surface brightness profiles consistent with low-level activity, equating to $\sim19$\% of our total sample. We note that 2020 RX133 is a Jupiter Trojan. When we considered the heliocentric distance range of the asteroids at the time they showed activity, this fraction increased to 33\% of the targets in the $3.16 \leq R_{h} \leq 4.56$ au region, and therefore it is possible to infer that at least $\sim30$ asteroids with $T_J \leq 3.05$ and in the $4.05 < a < 5.05$ au parameter space may potentially exhibit low-level activity. We also estimated  nuclear radii for the three active targets of $r_{n} = 1.8 \pm 0.2$ km, $r_{n} \leq 0.8$ km, and $r_{n} \leq 0.5$ km for (669525) 2012 XO144, 2011 WM183, and 2020 RX133 respectively. The median color index for the observed asteroids is $(g-r)_{PS1} = 0.52 \pm 0.13$, aligning with those expected for D-type asteroids. 
\end{abstract}

\section{Introduction} \label{sec:intro}
Jupiter-family comets (JFCs) originate from the Kuiper Belt, where slow orbital evolution near gravitational resonances can move them onto Neptune-crossing orbits \citep{2017ApJNesvorny}. Once within the orbits of the giant planets, these objects are then known as Centaurs. Several Centaurs have been shown to exhibit episodic activity, with outbursts releasing significant amounts of dust and gas between quiescent periods \citep{2009AJJewitt, 2010MNRASTrigo, 2016IcarMiles, 2019AJWong, 2023PSJDobson}. Centaurs can then continue their inward evolution towards the Sun and become JFCs where sublimation is typically dominated by H$_{2}$O within 3 au of the Sun \citep{1985A&AYamamoto, 2015SSRvCochran}. However, many JFCs have still been found to exhibit activity beyond 5 au \citep{2024PSJGillan,2025PSJGillan}, and even out to 7 au \citep{2013IcarKelley} where sublimation is likely driven by more volatile ices such as CO or CO$_{2}$. 

Our current evolutionary framework for short-period comets as they evolve inward is that of cometary nuclei that are dormant in the Kuiper Belt, gradually becoming more active as they move through the Centaur region and then finally showing a significant amount of activity (usually) once they have evolved into JFCs. The Tisserand parameter, $T_J$, is approximately conserved during close encounters with Jupiter, indicating that a body evolving inward with a $T_J < 3$ is likely a comet, while numerical simulations refine this to $T_J < 3.05$ \citep{2008ssbnGladman}. The vast majority of active comets indeed have $T_J < 3$. There are however many objects in such low-$T_J$ orbits that are classified as inert asteroids as there has been no reported activity. These objects are on comet-like orbits, known as ACOs (asteroids in cometary orbits; \citealt{2005AJFernandez, 2006AdSpRLicandro, 2018A&ALicandro, 2025ApJChandler}). Many of these objects may be inactive cometary nuclei as they typically possess similar spectra \citep{2018A&ALicandro, 2021MNRASSimion}. They also have similarly low albedos \citep{2014ApJKim}. It is possible that some of these objects are active, but have not yet been imaged to a sufficient depth at which low-level activity may be seen. 

Current wide-field surveys identify $\sim 15$ new JFCs each year. Activity is often detected via the 1.8 m PanSTARRS telescopes \citep{2016arXivChambers}, the 0.7 m -- 1.5 m Catalina Sky Survey telescopes \citep{2012DPSChristensen}, and 0.5 m ATLAS telescopes \citep{2018PASPTonry}, through images revealing a surrounding dust coma. With typical exposure times of only 30 seconds, giving limiting $5\sigma$ magnitudes of $\sim 19.5 - 21.5$, it is possible that many low-$T_J$ objects undergo weak sublimation that has not yet been detected by these surveys.

In this work, we observed asteroids with $T_J \leq 3.05$, and semi-major axis $4.05 < a < 5.05$ au, to search for dust comae associated with objects classified as asteroids. The known population in this region contained a mixture of 143 active comets and 153 asteroid-like objects at the time of observation. Hence, known comets form $\sim 50 \%$ of the small body population in this region, but it is possible that the fraction of active objects is higher. This study assesses the prevalence of faint dust comae in this subset of objects, providing insights into the extent of activity within the JFC population. Furthermore, this allows us to more accurately determine the fraction that are cometary nuclei currently undergoing sublimation and mass loss. 

\section{Target selection, observations and calibration } \label{sec:data}
\subsection{Target selection and observations }\label{target_selection}
The Isaac Newton Telescope (INT) is a 2.5 m optical telescope located at the Roque de los Muchachos Observatory on the island of La Palma in the Canary Islands. The imaging instrument that was used in this work was the Wide-Field Camera (WFC), which is located at the prime focus of the telescope and consists of four thinned EEV CCDs with pixel dimensions 2048 $\times$ 4100 arranged in an L-shape. The CCDs have a pixel size of 13.5 microns that corresponds to a pixel scale of \SI{0.33}{\arcsecond}/pixel. The wide-field nature of the camera was not required in this work, and therefore, only one of the CCDs was used to minimize the reduction workload. We used CCD 4 as this is on the optical axis of the telescope. We ensured our targets were near the center of this chip for each observation.

\begin{figure}
    \centering
    \epsscale{1.25}
    \plotone{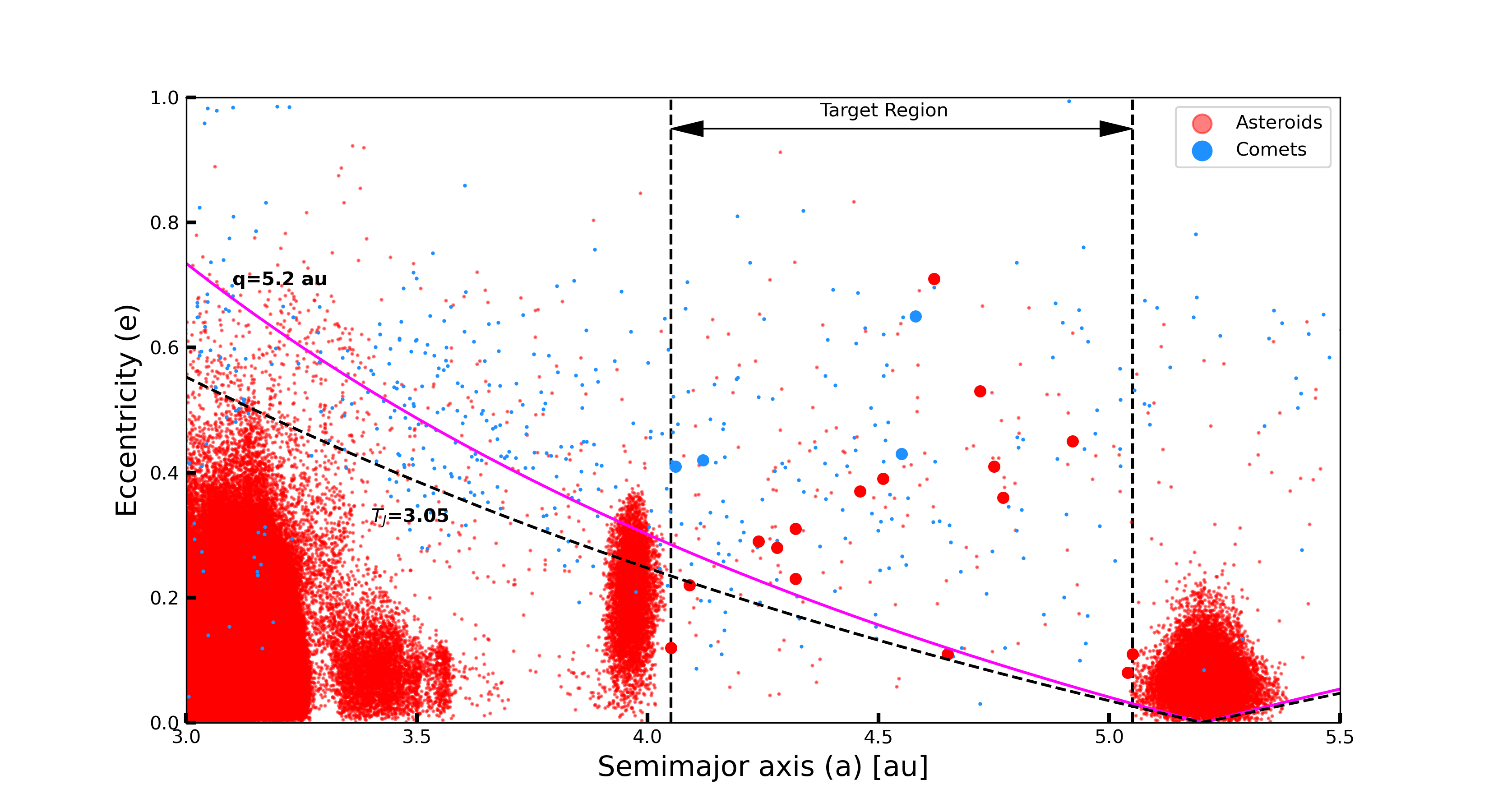}
    \caption{Orbital elements for asteroids/inert objects (red) and active comets (blue). The targets for this study are shown as large red dots, while the comparison comet targets are shown as large blue dots. Only the short-period comets 286P/Christensen, 291P/NEAT, 445P/Lemmon-PANSTARRS, and 446P/McNaught are included. The two long-period comets C/2022 J2 (Bok), C/2022 P3 (ZTF), and hyperbolic comet C/2022 U3 (Bok) are not shown due to their large semimajor axes. The lower black line, shows the Tisserand parameter with respect to Jupiter $T_\mathrm{J}$ $\leq$ 3.05 for 0$^\circ$ orbital inclination. Objects above the pink lines have Jupiter crossing orbits. The large asteroid populations (red) at 3.95 au and 5.20 au are the \emph{Hilda} and \emph{Trojan} families in stable resonance with Jupiter, respectively.}
    \label{fig:targets}
\end{figure}

As well as the above target criteria  in Section~\ref{sec:intro} of  $T_J \leq 3.05$ and semi-major axis $4.05 < a < 5.05$ au, we restricted ourselves to targets that had an observational arc $> 100$ days as a pragmatic filter to exclude the most poorly constrained orbits. We also rejected objects in the \emph{Thule} asteroid family, which occupies the 4:3 mean motion resonance with Jupiter, as these are stable and cannot have evolved there via the Centaur/JFC evolution. From this set of criteria, we selected all objects which were predicted to have apparent magnitudes $V \leq 22$ and observable from the INT during our observing run in November 2022. Figure~\ref{fig:targets} shows the target region and selected asteroids that were eventually observed in this study.

We observed our targets over a three-night observing window from November 15th to November 17th 2022 using the WFCSloanG ($g'$) and WFCSloanR ($r'$) filters with central wavelengths of $\lambda(g')=4846$\AA\ and $\lambda_(r') = 6240$\AA. These filters have the advantage of having sharp wavelength cut-offs between bands, allowing for precise color information as well as having well-defined standardized transmission curves. Although this observing run had no time lost to bad weather or clouds, the observing conditions were sub-optimal over the three nights with seeing ranging from $\sim1-2$\SI{}{\arcsecond}.

To optimise the observing time during the run, we prioritized asteroids most likely to show activity, categorising our target list into three groups. The primary targets, which were the highest priority, included asteroids with eccentricities $e >0.3$ and heliocentric distances $R_{h} < 4 $ au during the observing run. The primary targets were all observed using the Sloan $g'$ and $r'$ filters. The secondary targets consisted of asteroids with $R_{h} <4 $ au, regardless of eccentricity, and were observed with the same filters. Lastly, the tertiary targets were defined by $4.05 < a < 5.05 $ au, encompassing the remaining objects on our observing list without further criteria. The tertiary targets were only observed in the $r'$ filter.

We also observed several known comets, including long-period and short-period comets, with the only selection criterion being that they were observable with the INT during our observing window. These cometary observations served as comparative examples of active small bodies alongside our targets. Our final dataset is composed of sixteen different asteroids at different epochs, four short-period comets and three long-period comets. The full table of observations, including the observation times, heliocentric and geocentric distances, phase angles, number of exposures in each filter, and exposure time for each target, is presented in Table~\ref{tab:observations}. 

\startlongtable
\begin{deluxetable*}{lcccccc}
\tablecaption{Targets observed in this study by the INT. \textdagger\ - represents the comets in our sample and g and r refer to the Sloan $g'$ and $r'$ filters respectively. Exposure time refers to a single exposure.}
\label{tab:observations}
\tablehead{
\colhead{Name} & \colhead{Observation time} & \colhead{$R_{h}$} & \colhead{$\Delta$} & \colhead{$\alpha$} & \colhead{N$_{\textrm{exposures}} \times $Filter} & \colhead{Exposure Time} \\
 & \colhead{(UT)} & \colhead{(au)} & \colhead{(au)} & \colhead{(\textdegree)} & & \colhead{(s)}
}
\startdata
6144 Kondojiro (1994 EQ3) &                   2022 November 16 05:12-05:47 & 5.30 &  4.50 &        6.90 &                                    13 $\times$ r &      140 \\                  
        219844 (2002 CQ148) & 2022 November 18 02:42-03:15 & 4.71 &  3.84 &  6.43 & 12 $\times$ r & 150 \\     301964 (2000 EJ37) &  2022 November 16 03:30-04:07 & 5.34 &  4.47 &  5.49 & 17 $\times$ r & 110 \\
        2006 BV149 &                   2022 November 16 01:20-01:56 & 3.20 &  2.35 &       10.55 &                           12 $\times$ r & 150 \\                                 &                   2022 November 17 00:46-02:03 & 3.19 &  2.34 &       10.28 &        15 $\times$ r, 5 $\times$ g &      110 \\           
        434707 (2006 CH49) &                   2022 November 17 02:17-03:08 & 4.00 &  3.40 &       12.31 &        6 $\times$ r, 2 $\times$ g &      300 \\             371548 (2006 UY349) &                   2022 November 15 23:03-23:38 & 3.80 &  2.89 &        6.61 &             15 $\times$ r &      120 \\                                & 2022 November 16-17 23:33-00:42 & 3.81 &  2.90 &        6.88 &        18 $\times$ r, 6 $\times$ g &       95 \\           
        406520 (2007 VM224) &                   2022 November 16 19:28-20:35 & 3.75 &  3. 26 &       14.14 &     12 $\times$ r, 4 $\times$ g   &      180 \\               2010 LV121 &                   2022 November 17 03:16-04:18 & 3.88 &  3.33 &       13.20 &         12 $\times$ r, 4 $\times$ g &      150 \\                  
        2011 WM183 &                   2022 November 18 04:59-05:59 & 3.16 &  2.96 &       18.22 &         18 $\times$ r, 6 $\times$ g &       90 \\                     2012 UJ38 &                   2022 November 16 02:01-03:25 & 3.77 &  2.91 &        8.45 & 15 $\times$ r, 5 $\times$ g  & 150 \\                    
        (669525) 2012 XO144 &                   2022 November 15 20:02-20:45 & 3.37 &  2.68 &       13.63 &                                    13 $\times$ r &      175 \\ &2022 November 15-16 23:44-01:13&3.37&2.68&13.67&15 $\times$ r, 5 $\times$ g &175 \\             
        2014 WN504 &                   2022 November 15 22:25-22:57 & 3.35 &  2.83 &       15.72 &                                    10 $\times$ r &      185 \\                                 &                   2022 November 16 20:40-21:05 & 3.35 &  2.84 &       15.83 &                                     8 $\times$ r &      180 \\                                 &                   2022 November 17 19:36-21:03 & 3.34 &  2.85 &       15.93 &         15 $\times$ r, 5 $\times$ g &      185 \\            
        517594 (2014 WX199) &                   2022 November 17 23:23-23:55 & 4.27 &  3.62 &       10.81 &                                    10 $\times$ r &      180 \\               2015 MW21 &                   2022 November 18 03:57-04:56 & 3.33 &  3.08 &       17.19 &         18 $\times$ r, 6 $\times$ g &       90 \\                         2020 RX133 &                   2022 November 18 03:19-03:52 & 4.56 &  3.64 &        5.19 &                                    12 $\times$ r &      150 \\                      2021 LP4 &                   2022 November 15 20:50-21:29 & 2.94 &  2.43 &       18.23 &                                    22 $\times$ r &       80 \\                                 &                   2022 November 16 21:10-22-13 & 2.94 &  2.44 &       18.31 &         21 $\times$ r, 7 $\times$ g &       80 \\     
        286P/Christensen\textsuperscript{\textdagger} &                   2022 November 17 21:51-22:34 & 2.72 &  2.26 &       20.43 &                                    28 $\times$ r &       65 \\                     291P/NEAT\textsuperscript{\textdagger} &                   2022 November 17 21:30-21:47 & 2.82 &  2.29 &       19.01 &                                     8 $\times$ r &      120 \\   445P/Lemmon-PANSTARRS\textsuperscript{\textdagger} &                   2022 November 17 21:08-21:27 & 2.47 &  1.92 &       21.81 &                                    13 $\times$ r &       70 \\                 446P/McNaught\textsuperscript{\textdagger} &                   2022 November 17 22:37-23:18 & 2.39 &  1.51 &       13.74 &                                    16 $\times$ r &      110 \\ C/2022 J2 (Bok)\textsuperscript{\textdagger} &                   2022 November 16 06:07-06:10 & 1.85 &  1.47 &       32.27 &                                     5 $\times$ r &       20 \\            C/2022 P3 (ZTF)\textsuperscript{\textdagger} &                   2022 November 16 06:02-06:04 & 2.82 &  2.27 &       18.71 &                                     3 $\times$ r &       30 \\               C/2022 U3 (Bok)\textsuperscript{\textdagger} &                   2022 November 15 19:50-19:54 & 6.74 &  5.89 &        4.66 &                                     3 $\times$ r &      115 \\
\enddata
\end{deluxetable*}

For the asteroidal targets of this study, Table~\ref{tab:asteroid_targets} gives the orbital elements and each associated $T_J$. This subset of 16 asteroids covers a wide range of orbital parameters, perihelia, and $T_J$ values, giving a diverse and representative sample of the inactive objects within this region of parameter space.

\begin{table*}
\centering
\caption{Asteroids observed with the INT and their key orbital parameters.}
\begin{tabular}{lccccc}
\hline
Name & Eccentricity & Inclination & Semi-major axis & Perihelion distance & Tisserand Parameter \\
& & (\textdegree) & (au) & (au) & $T_J$\\
\hline
6144 Kondojiro (1994 EQ3) &              0.36 &             5.88 &           4.77 &          3.06 & 2.87 \\     
219844 (2002 CQ148) &              0.08 &             2.37 &                 5.04 &          4.64 & 2.99 \\      
301964 (2000 EJ37) &  0.71 &            10.07 &                 4.62 &                       1.36 & 2.43 \\         
2006 BV149 &              0.29 &            14.20 &                 4.24 &                   3.00 & 2.90 \\   
434707 (2006 CH49) &              0.28 &             7.66 &                 4.28 &           3.09 & 2.94 \\    
371548 (2006 UY349) &              0.12 &             3.34 &                 4.05 &          3.55 & 3.03 \\   
406520 (2007 VM224) &              0.22 &            13.14 &                 4.09 &          3.19 & 2.96 \\     
2010 LV121 &              0.45 &            19.83 &                 4.92 &                   2.71 & 2.69 \\        
2011 WM183 &              0.41 &            14.84 &                 4.75 &                   2.81 & 2.78 \\        
2012 UJ38 &              0.39 &            10.97 &                 4.51 &                    2.75 & 2.84 \\        
(669525) 2012 XO144 &              0.53 &            15.17 &                 4.72 &          2.24 & 2.66 \\        
2014 WN504 &              0.23 &             3.19 &                 4.32 &                   3.32 & 2.98 \\     
517594 (2014 WX199) &              0.11 &            17.63 &                 4.65 &          4.13 & 2.91 \\     
2015 MW21 &              0.31 &            15.50 &                 4.32 &                    2.99 & 2.87 \\         
2020 RX133 &              0.11 &             1.82 &                 5.05 &                   4.51 & 2.99 \\       
2021 LP4 &              0.37 &            19.73 &                 4.46 &                     2.79 & 2.79 \\
\hline
\end{tabular}
\label{tab:asteroid_targets}
\end{table*}

\subsection{Reduction and calibration}\label{calibration}

We reduced each of the INT images via bias and flat-field corrections. Bias images were taken at the start of each observing night and median combined into a master bias image which was subtracted from all flat-field and science images. The sky flats were taken in both the $g'$ and $r'$ filters at the start of each night (twilight) and similarly to the bias images were median combined. We also dithered the telescope during the twilight flat exposures to ensure any field stars disappeared when the frames were combined. The science images were divided by the flat frames of the same filter. Finally, the science images were astrometrically calibrated using the \emph{Astrometry.net} API \citep{2010AJLang} to match the pixels in the images to sky coordinates. 

Observing solar system objects using long, single exposures usually requires non-sidereal tracking or guiding. Instead, we imaged our targets using many short exposures with sidereal tracking, employing a `shift and stack' approach afterwards. For each of the targets we imaged, exposure times were limited to ensure a trailing of $\leq$\SI{0.5}{\arcsecond}. As the targets moved relative to background stars, we calculated the displacement between exposures and aligned all images. Sky background subtraction was then performed by measuring the local background in an annulus 30--50 pixels from the target. After the position-shift calculations and sky subtraction, we median combined the images to produce a single stacked image for each object. Finally, we photometrically calibrated all combined frames using the \emph{Panoramic Survey Telescope and Rapid Response System (Pan-STARRS) 1 Data Release 2 (PS1 DR2)} catalogue. Hence all photometry reported in this paper is in the PS1 system unless otherwise stated.

After visually verifying the targets in the combined images, the initial location of the target in each combined image was found using the predicted coordinates from \emph{JPL Horizons} \citep{giorginiJPLsOnLineSolar1996} and converting to pixel coordinates using \emph{Astropy wcs$\_$world2pix}\footnote{\url{https://docs.astropy.org/en/stable/api/astropy.wcs.WCS.html}}. The centroid of the target was measured by searching within a $10 \times 10$ pixel array around the predicted position and fitting a 2-D Gaussian profile using the \emph{Photutils centroid$\_$2dg}\footnote{\url{https://photutils.readthedocs.io/en/stable/api/photutils.centroids.centroid_2dg.html}} routine.

\section{Searching for cometary activity}
\subsection{Radial profile measurements}\label{radial_profiles}
An effective method to determine if there is any dust coma associated with a target asteroid is by measuring the radial profile of that asteroid and comparing this to the point spread function (PSF) of field stars. This technique has been employed in many studies to search for and quantify low-level activity {\it e.g.}  \citet{1987ApJJewitt}, \citet{2005MNRASLowry} and \citet{2017NaturMeech}. 
Active objects may show surface brightness profiles that exceed the stellar profile of the frame, depending on the level of activity, due to gas and dust being ejected from the nucleus into the coma. The stellar profile reflects the seeing conditions across all individual stacked images, as well as any errors introduced by the `shift and stack' method used for alignment. Any remaining difference between the stellar and target profiles can theoretically be attributed to the target not being a point source, indicating activity.

We measured the radial profiles of each asteroid in Table~\ref{tab:asteroid_targets} on each date they were observed. To measure the radial profile, we measured the flux in concentric circular apertures centered on the target using the \emph{Radial Profile} routine in \emph{Photutils} \citep{larry_bradley_2022_6825092}. For each of the target asteroids, we fitted a Moffat function to our radial profile. The 1-D Moffat function is given in Equation~\ref{eq:Moffat}, where $A$ is the amplitude of the model, $x_0$ is the x-position of the maximum value of the profile and $x$ is the radial position. The Moffat profile has two adjustable parameters, $\gamma$ and $\alpha$, which represent the core and wings of the profile, respectively, where $\alpha = 1$ and $\gamma = 2.5$ are commonly used as initial estimates for the Moffat profile fit \citep{2001MNRASTrujillo}.

\begin{equation}\label{eq:Moffat}
f(x) = A \left( 1 + \frac{(x-x_{0})^2}{\gamma^{2}} \right)^{-\alpha}
\end{equation}

\noindent For each Moffat profile fit, the FWHM was calculated for each asteroid using Equation~\ref{eq:FWHMMoffat}.

\begin{equation}\label{eq:FWHMMoffat}
FWHM = 2 \gamma \sqrt{2^{\frac{1}{\alpha}}-1}
\end{equation} 

To determine if activity was present on any of the asteroids we observed, we compared the Moffat profile of the asteroid to the radial profiles of field stars in each combined image. To retrieve the positions and magnitudes of field stars in the frame, the \emph{MAST API} was used to query the \emph{Pan-STARRS DR2} catalogue\footnote{\url{https://catalogs.mast.stsci.edu/panstarrs/}}. Our search for field stars was confined to a cone with a radius of \SI{320}{\arcsecond} centered on our target object. From this search, we extracted the brightest 120 non-saturated objects within the specified cone to form our stellar dataset. The brightest 120 non-saturated objects were selected to ensure well-measured stellar profiles with high signal-to-noise, while still retaining a large sample after outlier rejection and quality filtering; although these stars were generally brighter than the target, all were unsaturated and therefore suitable for PSF comparison. Each of these objects was then centroided using the same method as carried out for the target to ensure an accurate radial profile measurement. As the objects returned in the search could contain stars, galaxies and possible residual detector artifacts, we first separated stars and galaxies using the catalogued color $r(\textrm{PSFMag}) - r(\textrm{rKronMag}) < 0.05$\footnote{\url{https://outerspace.stsci.edu/display/PANSTARRS/How+to+separate+stars+and+galaxies}}.

The remaining profiles were subsequently normalized by their maximum intensity to facilitate inter-comparison. The median of the remaining profiles was then calculated and profiles that contained any value more than $2\sigma$ away from the median at any radii were removed. Two rounds of clipping were performed, with the median recalculated after the first iteration. This procedure ensured a clean sample of comparison stars for FWHM measurements and a more accurate, representative median stellar profile for PSF fitting. Typically, after applying these filtering criteria, the number of profiles was reduced from the initial set of 120 objects to $\sim 80 - 100$ stellar profiles, depending on the combined image. For the remaining stars, the median stellar profile was calculated and a Moffat function was fitted. A direct comparison with the Moffat profile fitted to the target radial profile was then made. 

As we used median-combined images, the sky background pixel values of the frame should tend to zero moving radially outward from a source. A correction was applied to the combined images where the background was not exactly zero. This was achieved by measuring the local background in our combined image and subtracting this value from each pixel value to get a background value of zero. Finally, to measure any excess flux, we converted the radial flux measurements to surface brightness in magnitudes/arcsecond$^{2}$. 

\subsection{FWHM as an indicator of activity}\label{subsec:fwhm_compare}
\begin{figure}[h]
    \centering
    \epsscale{1.25}
    \plotone{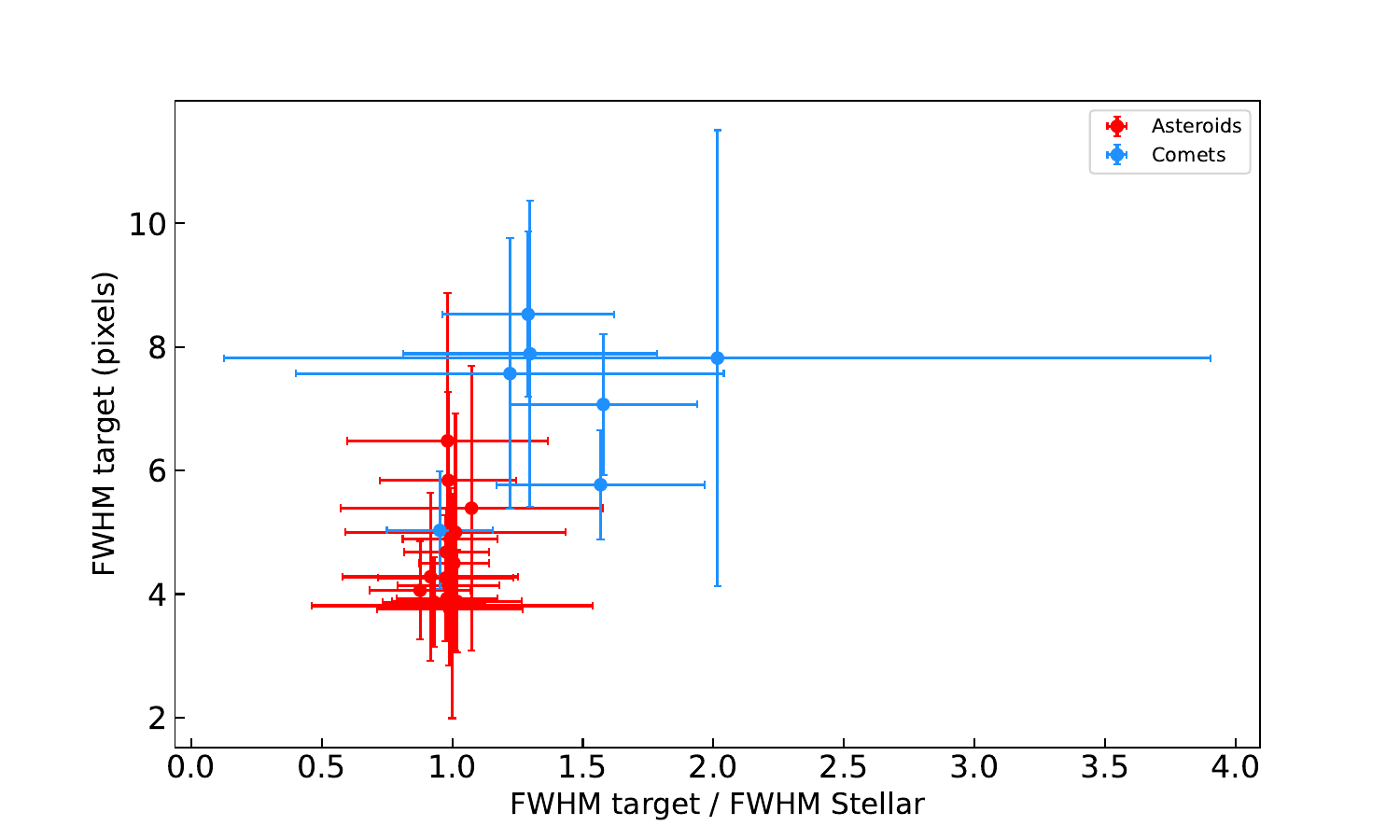}
    \caption{FWHM of the Moffat profiles of the asteroids and known comets in our dataset. The asteroids are shown in red and the comets are shown in blue.}
    \label{fig:FWHM}
\end{figure}

We investigated the relationship between the full width at half maximum (FWHM) of the Moffat fit for each asteroid and that of the median stellar profile in the same field, to assess whether activity could be identified through a more straightforward comparison. Figure~\ref{fig:FWHM} presents this comparison for both asteroids and known comets. As expected, two distinct populations are evident in the figure, corresponding to the comets and asteroids. The active comets generally exhibit a larger FWHM, which differs significantly from the FWHM of the median stellar profile in the frame. Asteroids show a FWHM comparable to that of the stellar profile. There is one comet -- 446P/McNaught -- that exhibits a FWHM comparable to that of the median frame stellar profile. Upon inspection, we found that this comet was apparently inactive during our observations, displaying a PSF similar to that of the field stars.

Overall, Figure~\ref{fig:FWHM} shows that this method is effective for distinguishing clearly active comets from inert objects and may serve as a useful quick-reference tool. However, it lacks the sensitivity to detect low-level or subtle activity. Small deviations from the expected 1:1 ratio are likely caused by the limitations of purely numerical FWHM comparisons, as subtle low-level activity may not be captured by the Moffat fit to the target profile. As such, further analysis using techniques such as the radial profile analysis above is required to detect minimal activity.

\subsection{Active targets}\label{activity_int}
From our dataset of 16 different asteroids, two of them (6144 Kondojiro (1994 EQ3) and 2014 WN504) passed too close to stars to allow accurate profile measurements beyond \SI{2}{\arcsecond}. However it was clear from visual inspection that they did not possess obvious dust comae, in agreement with the FWHM analysis above. For the remaining 14 targets, we obtained 17 surface brightness profiles, as some were observed on multiple nights. To ensure accurate measurements, we verified that none of the asteroids appeared near background stars or galaxies in any of the images, thereby avoiding flux contamination. This approach allowed us to attribute any observed activity unambiguously to the asteroid itself in each case. The surface brightness profiles for asteroids 2011 WM183, (669525) 2012 XO144, and 2020 RX133 exhibited a significant increase in surface brightness above the median stellar profile, consistent with low-level cometary activity. For each of these three potentially active targets, we also measured the surface brightness profile gradients using the relationship described in Equation~\ref{eq:SB_gradient}.

\begin{equation}\label{eq:SB_gradient}
m = \frac{d\log F(\rho)}{d\log\rho}
\end{equation}

Here, $m$ is the activity gradient, and $F(\rho)$ is the observed flux at $\rho$. For a steady-state coma model, we would expect a surface brightness gradient of $m = -1$ as the volume density of the isotropically emitted dust grains is $N_{d} \propto 1/\rho^{2}$. Any of the target asteroids that have a profile gradient $-1.5\leq m < -1$ are most likely due to the radiation pressure acting on the dust grains \citep{1987ApJJewitt, 2008MNRASSnodgrass}. An activity gradient of $m<-1.5$ may indicate the presence of sublimating grains \citep{2005MNRASLowry}.

\subsubsection{2011 WM183}
\begin{figure}[h]
    \centering
    \epsscale{1.25}
    \plotone{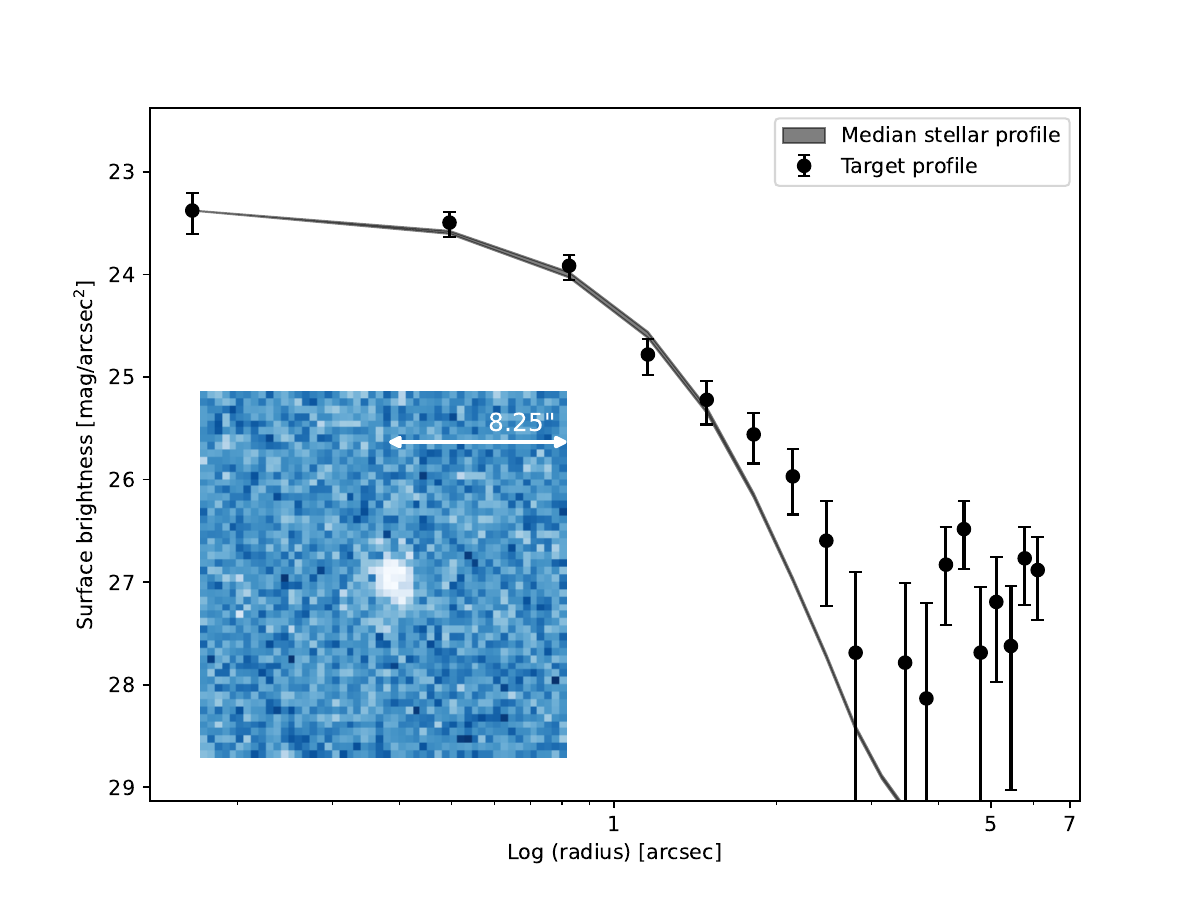}
    \caption{The median stellar profile and associated standard deviation are shown by the shaded gray area. The inset co-added image of 2011 WM183 composed of $18\times90$ second exposures on 2022 November 18. The equivalent exposure of the combined frame is 27 minutes. The surface brightness profile shows that 2011 WM183 is likely active as the measured profile and uncertainties exceed the stellar profile.}
    \label{fig:SB_2011WM183}
\end{figure}

We observed 2011 WM183 on 2022 November 18 at $R_{h} = 3.16$ au. The surface brightness profile of this asteroid, shown in Figure~\ref{fig:SB_2011WM183}, deviates from the median stellar profile at $\sim \SI{2}{\arcsecond}$ and continues to show an increase in surface brightness above the stellar profile. Although the uncertainties are large for the majority of the surface brightness measurements, there is clear separation from the median stellar PSF at $\rho \geq$ $\SI{2}{\arcsecond}$.

To measure the activity gradient, we used a least squares fit method in the region where there was a clear separation between the target and the stellar profile. For 2011 WM183, we used the outer 12 radii measurements between $\rho\simeq \SI{2.5}{\arcsecond}$ and $\rho \simeq \SI{6.5}{\arcsecond}$. The activity gradient of 2011 WM183 is $m = -1.8 \pm 0.7$. This is consistent with a radiation pressure dominated dust coma at the $1\sigma$ level, or a simple expanding dust coma at the $2\sigma$ level. 

\subsubsection{(669525) 2012 XO144}
\begin{figure}[h]
    \centering
    \epsscale{1.25}
    \plotone{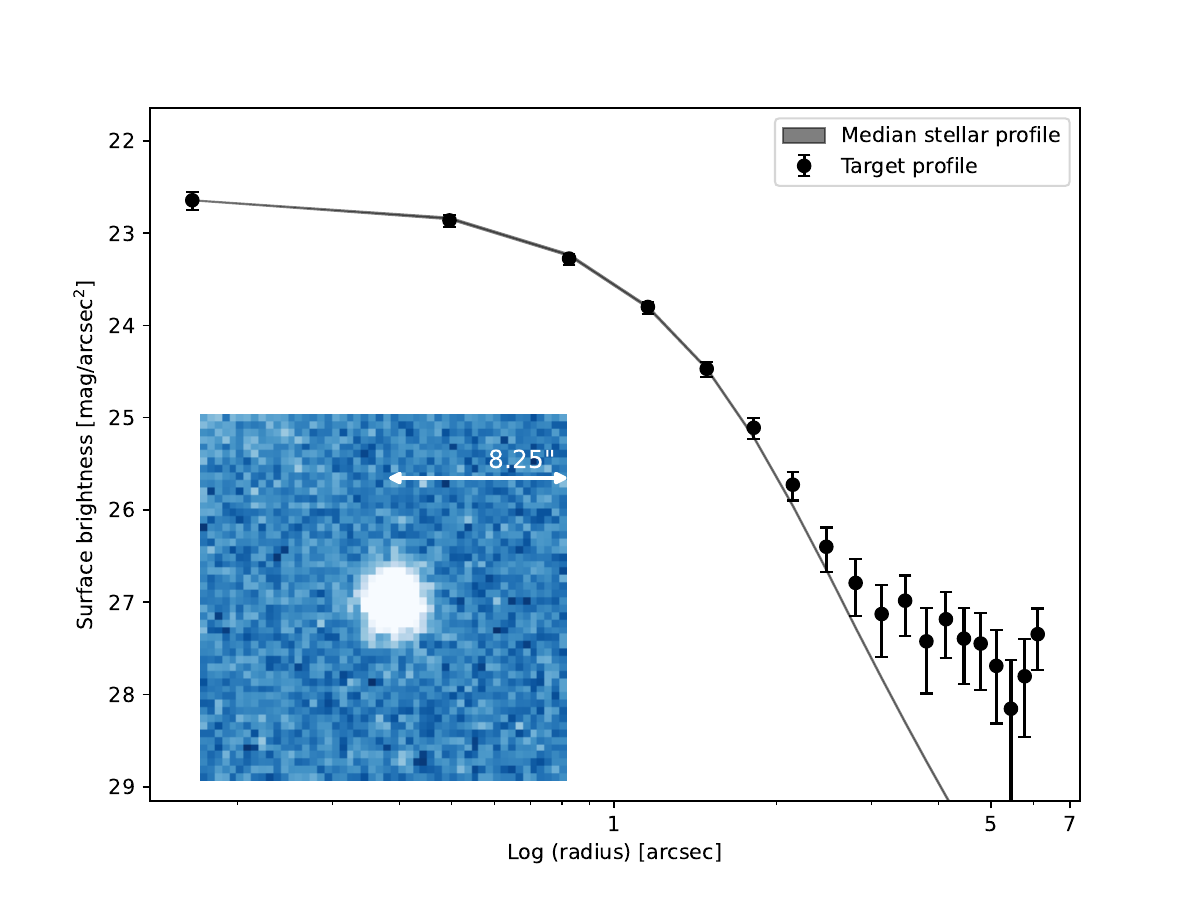}
    \caption{The median stellar profile and associated standard deviation are shown by the shaded gray area. The inset co-added image of (669525) 2012 XO144 composed of $28\times175$ second exposures on 2022 November 15. The equivalent exposure of the combined frame is 1.36 hours. The surface brightness profile shows that (669525) 2012 XO144 is likely active as the measured profile and uncertainties exceed the stellar profile.}
    \label{fig:SB_2012XO144}
\end{figure}

We observed (669525) 2012 XO144 on 2022 November 15 at a heliocentric distance of $R_{h} = 3.37$ au. The profile of this target, shown in Figure~\ref{fig:SB_2012XO144}, follows the stellar profile until $\sim \SI{3.5}{\arcsecond}$ from the center of the target where the separation of the two profiles becomes apparent. The active region of the surface brightness profile is clearly separated from the stellar profile by $> 2\sigma$ for all surface brightness measurements.

We measured the surface brightness gradient over the outer 9 radii measurements where $m$ decreases monotonically with increasing $\rho$ between $\rho \simeq \SI{3.5}{\arcsecond}$ and $\rho \simeq \SI{6.5}{\arcsecond}$. The surface brightness slope of (669525) 2012 XO144 is $m = -2.2 \pm 1.3$. Due to the large uncertainty, it is difficult to determine if the measurement is consistent with a simple expanding dust coma, or a radiation pressure dominated dust coma.

\subsubsection{2020 RX133} 
\begin{figure}[h]
    \centering
    \epsscale{1.25}
    \plotone{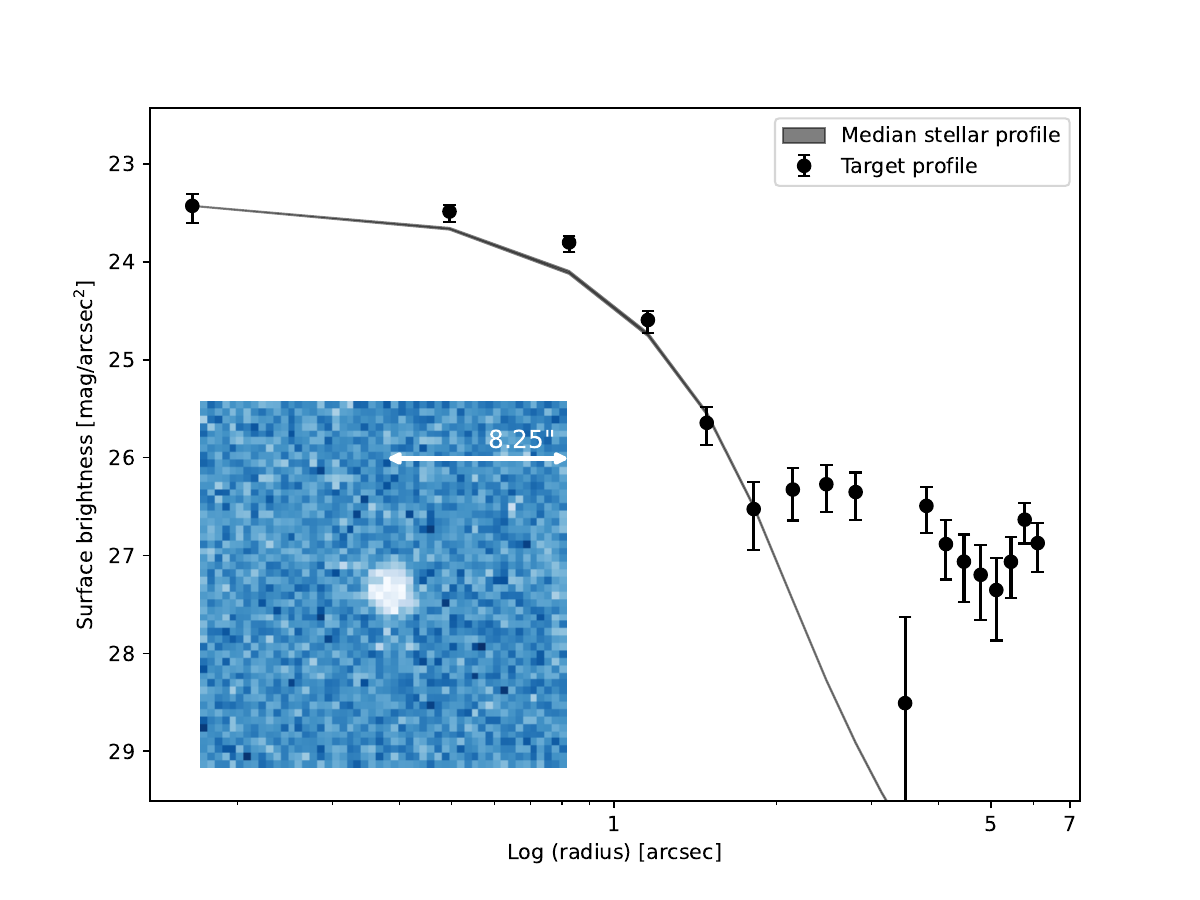}
    \caption{The median stellar profile and associated standard deviation are shown by the shaded gray area. The inset co-added image of 2020 RX133 composed of $12\times150$ second exposures on 2022 November 18. The equivalent exposure of the combined frame is 30 minutes. The surface brightness profile shows that 2020 RX133 is likely active as the measured profile and uncertainties exceed the stellar profile.}
    \label{fig:SB_2020RX133}
\end{figure}

We observed 2020 RX133 on 2022 November 18 at $R_{h} = 4.56$ au. The profile of this target, shown in Figure~\ref{fig:SB_2020RX133}, begins to deviate from the stellar profile at $\sim$ \SI{2.5}{\arcsecond}. The shape of the active region of the target profile appears flatter than the two previous examples, suggesting the activity should be more obvious and apparent, however from the inset image this is not the case.

We measured the activity gradient over the outer 11 radii between $\rho \simeq \SI{2.5}{\arcsecond}$ and $\rho \simeq \SI{6.5}{\arcsecond}$, again being careful to restrict the activity gradient measurements to where the stellar profile and the target profile have clearly deviated. The activity slope of 2020 RX133 is $m = -1.6 \pm 0.5$. This profile has the shallowest gradient of the three active targets, which matches the surface brightness profile, and is consistent with a radiation pressure-dominated dust coma at the $1\sigma$ level, or a simple expanding dust coma at the $2\sigma$ level. 

\subsection{Comparison with cometary activity profiles}
To provide context for the low-level activity detected in the three aforementioned asteroids, we compared their surface brightness profiles with that of a known comet - 446P/McNaught. This JFC was the least active of our comparison comets at the time of observing. This JFC has a perihelion distance of $q=1.62$ au, an eccentricity of $e=0.65$, a semi-major axis of $a=4.58$ au, and an orbital inclination of $i=16.50^\circ$. At the time of observation, 446P/McNaught was post-perihelion at an $R_{h} = 2.39$ au and appeared as a point source in the combined image. The surface brightness profile of this JFC shows a deviation from the median stellar profile at $\sim\SI{2.5}{\arcsecond}$; however, it is unclear from this profile if the comet is active at the time of observing or whether this is residual dust from a previous period of activity.

\begin{figure}[h]
    \centering
    \epsscale{1.25}
    \plotone{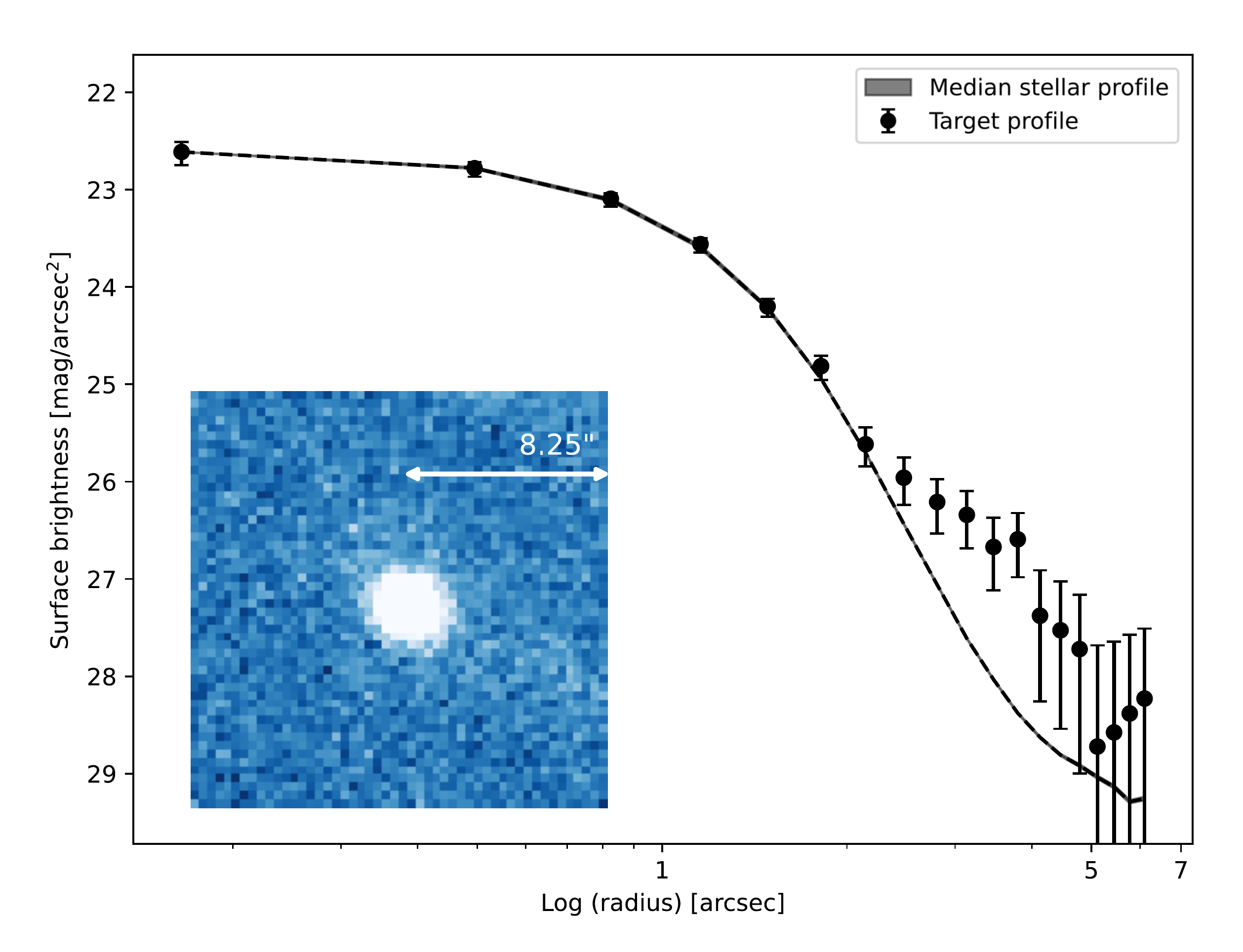}
    \caption{The median stellar profile and associated standard deviation are shown by the shaded gray area. The inset co-added image of 446P/McNaught composed of $16\times110$ second exposures on 2022 November 17. The equivalent exposure of the combined frame is 29 minutes. The surface brightness profile shows that 446P/McNaught was likely inactive (or very subtly active) as the measured profile and uncertainties do not exceed the stellar profile at $> 2\sigma$.}
    \label{fig:SB_446P}
\end{figure}

When compared to the active asteroids in our sample, 446P/McNaught exhibits a surface brightness profile that remains close to the median stellar profile, with no separation at larger radii, possibly suggesting a compact coma. The active asteroids (2011 WM183, (669525) 2012 XO144, and 2020 RX133), by contrast, show larger deviations from the stellar profile. This comparison demonstrates that our potentially active asteroids may exhibit more pronounced coma-like behavior, comparable to-or even exceeding-that of a possibly weakly active JFC.

\subsection{Archival Image Search} \label{susbec:archivalImageSearch}
We carried out a search for archival images of these objects via four primary avenues (1) the Hunting for Activity in Repositories with Vetting-Enhanced Search Techniques (HARVEST; \citealt{chandlerChasingTailsActive2022a,chandlerActiveAsteroidsCitizen2024}), which is focused on Dark Energy Camera (DECam) data but also includes the Canada-France-Hawaii Telescope (CFHT) MegaPrime dataset, (2) the Canadian Astronomical Data Centre (CADC) Solar System Object Information Search\footnote{\url{https://www.cadc-ccda.hia-iha.nrc-cnrc.gc.ca/en/ssois/}} (SSOIS; \citealt{gwynSSOSMovingObjectImage2012}), which for this search yielded results from DECam, MegaPrime, OmegaCam, Pan-STARRS~1, and SkyMapper, and (3) the NASA/CalTech Infrared Science Archive (IRSA) Moving Object Search Tool (MOST)\footnote{\url{https://irsa.ipac.caltech.edu}} of Zwicky Transient Facility (ZTF) and Palomar Transient Factory (PTF) archives, and (4) searching our own observing records for past data.

In total, our public archive search yielded 3,817 images with fields of view that contained objects from Table \ref{tab:asteroid_targets}. Our internal search revealed observations of two objects, 2006 CH49 from UT 2024 April 11 and 2014 WN504 from UT 2023 October 15 and UT 2024 January 5), all via observations with the Apache Point Observatory (APO) 3.5~m telescope and the Astrophysical Research Consortium Telescope Imaging Camera (ARCTIC). We examined each image individually to search for visible evidence of activity. We conducted this search by-eye as the methods described in, e.g., Section \ref{subsec:fwhm_compare}, have not been adapted to work with the myriad of archival instruments we searched; such an adaptation would take a major effort, and is beyond the scope of this work. We found no visible evidence of previous activity in the public and internal archival data we searched.

\subsection{Observed magnitudes}\label{sec:magnitudes}
For each of the comets and asteroids observed, we performed aperture photometry to measure their fluxes and derive corresponding magnitudes. The $r'$-band magnitudes of our targets are listed in Table~\ref{tab:aper_phot_int}. The magnitudes for known comets were measured through apertures of radii $10^4$ km at the comets. Interestingly, the two faintest asteroidal targets were two that displayed non-stellar surface brightness profiles (Figures~\ref{fig:SB_2011WM183} and \ref{fig:SB_2020RX133}). 

\begin{table}[h]
\centering
\begin{tabular}{lc}
\hline
Target & Apparent $r'$-band magnitude  \\
\hline
\textbf{Night 1} \\
2006 BV149 & $20.68\pm0.04$ \\
2012 UJ38 & $21.91\pm0.05$ \\
(669525) 2012 XO144 & $21.19\pm0.03$ \\
2014 WN504 & $21.37\pm0.04$ \\
2021 LP4 & $20.96\pm0.04$ \\
6144 Kondojiro (1994 EQ3) & $19.41\pm0.03$ \\
301964 (2000 EJ37) & $21.01\pm0.04$ \\
371548 (2006 UY349) & $21.42\pm0.05$ \\
C/2022 J2 (Bok) & $19.66\pm0.1$ \\
C/2022 U3 (Bok) & $19.98\pm0.04$ \\
C/2022 P3 (ZTF) & $18.99\pm0.05$ \\
\hline
\textbf{Night 2} \\
2006 BV149 & $20.51\pm0.05$ \\
2010 LV121 & $21.55\pm0.05$ \\
2014 WN504 & $21.25\pm0.04$ \\
2021 LP4 & $21.05\pm0.05$ \\
371548 (2006 UY349) & $21.35\pm0.04$ \\
406520 (2007 VM224) & $21.61\pm0.05$ \\
434707 (2006 CH49) & $21.24\pm0.04$ \\
\hline
\textbf{Night 3} \\
286P/Christensen & $20.10\pm0.03$ \\
291P/NEAT & $19.37\pm0.05$ \\
445P/Lemmon-PANSTARRS & $20.38\pm0.05$ \\
446P/McNaught & $20.96\pm0.03$ \\
2011 WM183 & $22.02\pm0.08$ \\ 
2014 WN504 & $21.51\pm0.05$ \\
2015 MW21 & $21.93\pm0.06$ \\
2020 RX133 & $22.07\pm0.09$ \\
219844 (2002 CQ148) & $20.27\pm0.06$ \\
517594 (2014 WX199) & $21.19\pm0.04$ \\
\hline
\end{tabular}
\caption{Calibrated $r'$-band magnitudes for all objects observed with the WFC on the INT during the observing run.}
\label{tab:aper_phot_int}
\end{table}

\begin{figure}[h]
    \centering
    \epsscale{1.25}
    \plotone{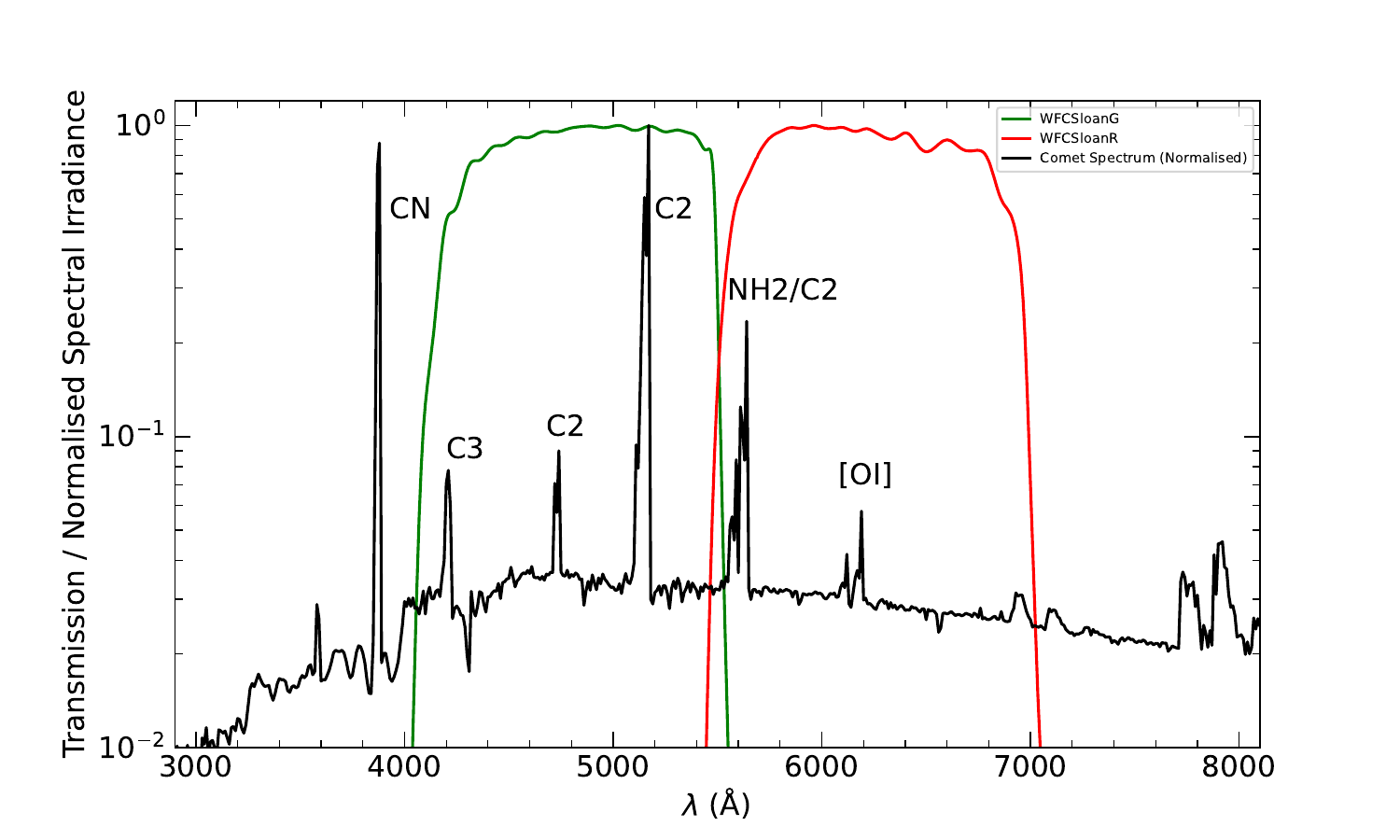}
    \caption{The transmission curves for the Sloan $g'$ and $r'$ filters are shown in green and red respectively. A standard comet spectrum is plotted in black with the molecular emissions labeled above each feature in the spectrum.}
    \label{fig:INT_and_spectra}
\end{figure}

As we used the Sloan $g'$ and $r'$ bands in our observations, we show the transmission curves in Figure~\ref{fig:INT_and_spectra} with a template comet spectrum generated by NASA's Planetary Spectrum Generator \citep{2018JQSRTVillanueva}\footnote{\url{https://psg.gsfc.nasa.gov}}. For very active comets, the figure highlights the dominance of gas emission in the C$_2$ band within the wavelength coverage of the Sloan $g'$ filter. The $g'$ filter also contains some minor C$_3$ emission. The $r'$ filter is potentially contaminated by C$_2$, and weaker emission bands of NH$_2$ and [OI]. However, distant low-activity comets rarely show detectable gas emission at these wavelengths, and we assume that any coma in our $r'$-band images are dominated by dust.

\subsection{Coma magnitudes and $Af\rho$}
We calculated the total coma for our three active targets and the known comets, by measuring the surface brightness at $r = \SI{5}{\arcsecond}$ from the object. From our measured surface brightness profiles, the coma at this distance should dominate the observed flux. We then used the relationship from \citet{1984IcarJewitt} to estimate the total coma contribution in our $r'$-band measurements as shown in Equation~\ref{eq:SB}.

\begin{equation}\label{eq:SB}
m_{coma} = \Sigma (r) -2.5 \log (2 \pi r^{2}) 
\end{equation}

\noindent In this equation, $m_{coma}$ represents the integrated magnitude of an assumed steady state coma within a circular radius $r$ in arcseconds. $\Sigma(r)$ is the surface brightness measured at the same radius expressed in magnitudes/$\textrm{arcsecond}^{2}$.

We also estimated the relative dust production rates of the active targets using the $Af\rho$ parameter \citep{1984AJAHearn} and corrected the phase angle to $0^\circ$ using the Schleicher-Marcus phase function \citep{2010Schleicher}. For the $A(0^\circ)f\rho$ measurements of the asteroids, we did not use the conventional $\rho = 10,000$ km aperture as is typical for JFC studies \citep{2013A&ASnodgrass, 2016MNRASBoehnhardt, 2022IcarBorysenko, 2023PASJLin, 2024PSJGillan}, as this was larger than the coma detected for our three active targets, but instead used a \SI{5}{\arcsecond} radius aperture. At their observed geocentric distances,  this was equivalent to $\rho = 9700$ km for (669525) 2012 XO144,  $\rho = 10,700$ km for  2011 WM183, and : $\rho = 13,200$ km for 2020 RX133. The corresponding, measured $A(0^\circ)f\rho$ values were $A(0^\circ)f\rho = 3.0\pm1.0$ cm for (669525) 2012 XO144, $A(0^\circ)f\rho = 4.1\pm2.1$ cm for 2011 WM183, and $A(0^\circ)f\rho = 7.4\pm2.1$ cm for 2020 RX133. As expected, these values are similar to some of the lowest values observed for cataloged comets \citep{2024PSJGillan, 2025PSJGillan}. Comparison to main belt comets \citep{2006SciHsieh, 2015IcarHsieh} however, may be more appropriate than comparison with the typically more active JFCs. For example, \citet{2013ApJHsieh} measured an $A(0^\circ)f\rho$ ranging from 6--13 cm within a \SI{5}{\arcsecond} aperture for main-belt comet 358P/PANSTARRS. \citet{2021ApJHsieh} and \citet{2022MNRASNovakovic} also measured $A(0^\circ)f\rho$ ranging from 10--25 cm for the main-belt comet 433P/(248370) 2005 QN173. Both of these studies provide more suitable comparisons to our measurements. However, we note that $Af\rho$ is poorly defined for very low-activity objects, as it assumes that all of the flux comes from the coma. This is not the case when the signal from the nucleus contributes to a significant fraction of the flux, which may be true here.

\begin{table}[h]
\centering
\begin{tabular}{lc}
\hline
 & $m_{\mathrm{coma}}$  \\
\hline
\textbf{ACTIVE TARGETS} \\
(669525) 2012 XO144 & $22.1\pm0.4$  \\
2011 WM183 & $21.9\pm0.6$  \\ 
2020 RX133 & $21.8\pm0.3$  \\
\hline
\textbf{COMETS} \\
C/2022 J2 (Bok) & $18.8\pm0.2$ \\
C/2022 U3 (Bok) & $20.5\pm0.2$  \\
C/2022 P3 (ZTF) & $19.1\pm0.2$  \\ 
286P/Christensen & $21.0\pm0.4$  \\
291P/NEAT & $19.6\pm0.1$  \\
445P/Lemmon-PANSTARRS & $20.2\pm0.2$  \\
446P/McNaught & $22.9\pm0.8$  \\
\hline
\end{tabular}
\caption{Coma magnitudes for the active targets and known comets.}
\label{tab:coma_mag}
\end{table}

\subsection{Nuclear magnitudes and radii estimates}
To measure the apparent magnitudes of our active targets, we took the total flux from the active object and subtracted the contribution from the coma, leaving only the flux from the nucleus. Working in terms of magnitudes, shown in Tables \ref{tab:aper_phot_int} and \ref{tab:coma_mag}, we used the relationship in Equation~\ref{eq:mag_nuc_int} 

\begin{equation}\label{eq:mag_nuc_int}
m_{nucleus} = m_{total} - 2.5\log_{10}[1 - 10^{0.4(m_{total} - m_{coma})}]
\end{equation}

\noindent where $m_{nucleus}$ is the magnitude contribution from the nucleus only, $m_{total}$ is the total magnitude from the object obtained from the aperture photometry, and $m_{coma}$ is the magnitude contribution from the coma. We then calculated the absolute magnitude, \emph{H}, at 1 au from the Sun and Earth at 0{\textdegree} phase angle ($\alpha$) using a linear phase coefficient common in cometary studies of $\beta$ = 0.035 \citep{2005MNRASLowry, 2009A&ALamy}.

\begin{equation}\label{eq:Hmag}
H = m_{nucleus}-5\log_{10}(r_{H}{\Delta})-{\beta}{\alpha}
\end{equation}

For asteroids, the H-G magnitude or HG12 systems are commonly used \citep{2010IcarMuinonen}. These phase functions include an `opposition surge', which causes an increase in brightness at small phase angles and a non-linear response at large phase angles. There are no observed opposition surges in cometary nuclei phase curves \citep{2024cometsIIIKnight}. Our use of a linear phase function is justified as none of our observations are at very high or very low phase angles. We subsequently calculated the radius of the nuclei using the relationship between the absolute magnitude and the radius of the cometary nucleus from the relationship of \citep{1916ApJRussell}:

\begin{equation}\label{eq:rn_ast}
A_{r}r_{n}^{2} = 2.25\times10^{22} \, 10^{0.4(m_{\odot} - H)}
\end{equation}

\noindent Here, $A_{r}$ is the geometric albedo, $r_{n}$ is the nuclear radii in metres, and $m_{\odot}$ is the apparent magnitude of the Sun in the PS1 $r'$ filter of $M_{\odot}=-26.93$ \citep{2018ApJSWillmer}. The resulting absolute magnitudes and corresponding radii estimates are shown for the three active targets in Table~\ref{tab:Rn_ast}.

\begin{table}[h]
\centering
\begin{tabular}{lcc}
\hline
 & $H$ & $r_{n}$\\
 & & (km) \\
\hline
\textbf{ACTIVE TARGETS} \\
(669525) 2012 XO144 & $16.2\pm 0.3$ & $1.8\pm 0.2$ \\
2011 WM183 & $\geq 17.8$ & $\leq 0.8$ \\ 
2020 RX133 & $\geq 19.2$ & $\leq0.5$ \\
\hline
\end{tabular}
\caption{Absolute magnitudes and nuclear radii measurements for the active targets. We have assumed a geometric albedo of $A_{r}$ = 0.04 and a phase coefficient of $\beta$ = 0.035. 2011 WM183 and 2020 RX133 are 1$\sigma$ upper limits.}
\label{tab:Rn_ast}
\end{table}

For 2011 WM183 and 2020 RX133, we measured 1$\sigma$ upper limits as the estimated total coma magnitude was brighter than the measured total magnitude from the aperture photometry {\it i.e.} for 2011 WM183, we used $m_{coma}$ = 21.9 $+$ 0.6 = 22.6. 

From \emph{JPL Horizons} \citep{giorginiJPLsOnLineSolar1996}, (669525) 2012 XO144 and 2011 WM183 are classed as asteroids and 2020 RX133 is classed as a Jupiter Trojan
(see Section~\ref{trojan} for further discussion of this object).
It is therefore more likely that (669525) 2012 XO144 and 2011 WM183 evolved inwards to their current orbits and 2020 RX133 has not.
The nuclei of the JFCs range from a diameters of $\simeq$ 0.3 km up to more than 10 km \citep{2011MNRASSnodgrass, 2017MNRASKokotanekova, 2024cometsIIIKnight}. If (669525) 2012 XO144 and 2011 WM183 are in the process of evolving inwards and are in transition from Centaur to JFC, they fall within the measured size range for JFC nuclei.

\subsection{Color Indices}
We observed each of the primary and secondary targets (Section~\ref{target_selection}) in the order $r'-g'-r'$. This sequence enabled us to interpolate between the two $r'$-band observations to obtain an equivalent $r'$-band magnitude at the same time as the $g'$-band observation. To perform this interpolation, we combined the $r'$-band images (pre-$g'$) into a single stacked image, and similarly for the $r'$-band measurements taken after the $g'$ band images. We also combined the $g'$-band images. This resulted in three combined images suitable for photometry. With two $r_{PS1}$-band magnitudes obtained on either side of the $g_{PS1}$-band magnitude, we interpolated to determine the equivalent $r_{PS1}$-band magnitude at the time of the $g'$-band observation. From this, we calculated the $(g-r)_{PS1}$ color indices shown in Table~\ref{tab:colors}.

D-type asteroids are known to dominate the bodies found in the outer edge of the main asteroid belt and within Jupiter's Trojan regions \citep{2014NaturDeMeo, 2019NatAsFujiya}. Dynamical models based on the Nice model \citep{2005NaturTsiganis} suggest that D-types were formed in the outer solar system beyond Neptune’s orbit. They were then transported inwards during the giant planet migration \citep{2014NaturDeMeo, 2023AcAauRibeiro}, a migration caused by the dynamic evolution of the giant planets \citep{2009NaturLevison}. \citet{2003AJFernandez} also suggests that the surfaces of D-type asteroids are more similar to active and post-active comets, and not like pre-active comets (Centaurs). 

\begin{table}[h]
\centering
\begin{tabular}{lc}
\hline
Target & $(g-r)_\mathrm{PS1}$  \\
\hline
2012 UJ38 & $0.58\pm0.11$ \\
(669525) 2012 XO144 & $0.43\pm0.06$ \\
2006 BV149 & $0.52\pm0.06$ \\
2010 LV121 & $0.52\pm0.08$ \\
2021 LP4 & $0.49\pm0.08$ \\
371548 (2006 UY349) & $0.31\pm0.07$ \\
406520 (2007 VM224) & $0.55\pm0.08$ \\
434707 (2006 CH49) & $0.60\pm0.08$ \\
2011 WM183 & $0.28\pm0.17$ \\ 
2014 WN504 & $0.56\pm0.10$ \\
2015 MW21 & $0.22\pm0.13$ \\ \hline
\textbf{Median color} & $\mathbf{0.52\pm0.13}$ \\
\hline
\end{tabular}
\caption{$(g-r)_\mathrm{PS1}$ colors for asteroids in this study.}
\label{tab:colors}
\end{table}

The median $(g-r)_{PS1}$ color index and standard deviation for our asteroid sample is $(g-r)_{PS1} = 0.52 \pm 0.13$, equivalent to $(g'-r')_\mathrm{SDSS}  = 0.44\pm 0.13$ using the photometric transformations of 
\cite{2012TonryApJ}. The large standard deviation reflects the significant variation of color among the asteroids studied. Looking at the two color-segregrated populations of Trojans, \citet{2014AJWong} measured $(g'-r')_{SDSS}  = 0.62$ for red objects, and $(g'-r')_{SDSS}  =  0.52$ for less red objects. If we consider the median of these two measurements, $(g'-r')_{SDSS} = 0.57$ which is equivalent to $(g-r)_{PS1} = 0.48$, the median color of our objects agrees with that expected for D-type asteroids to within $1\sigma$. Notably three asteroids, (371548) 2006 UY349, 2011 WM183 and 2015 MW21, exhibit significantly lower $g-r$ values. However all three agree with the color of less-red Trojans within $2\sigma$.

In our study, the three active targets: (669525) 2012 XO144, 2011 WM183 and 2020 RX133, have $(g-r)_{PS1}$ colors of $0.43\pm0.06$, $0.28\pm0.17$, and no measurement for 2020 RX133, as it was only observed in the $r'$-band. For (669525) 2012 XO144, $(g-r)_{PS1}$ is consistent with the sample median within 1$\sigma$. 2011 WM183's color is also consistent within 1$\sigma$, however the measurement carries a large uncertainty. The larger uncertainty for this asteroid stems from larger photometric errors in both the $g'$ and $r'$ bands, where each combined stacked image for the color analysis consisted of 6 $\times$ 90 second exposures, yielding an effective exposure time of 9 minutes per image with a lower signal to noise than the rest of the sample.

\section{Discussion}\label{sec:discussion}
\subsection{The active asteroidal targets}\label{active_targets_discussion}
From our sample of 16 targets observed over the 3 nights, we measured FWHM for all of them and surface brightness profiles for 14 of them. Three showed surface brightness profiles that could be interpreted as evidence of dust comae, equating to $\sim19$\% of our total sample. The three active targets were at heliocentric distances between 3.16 $\leq$ $R_{h} \leq 4.56$ au. Of the 16 targets, 9 were observed within this heliocentric distance range, equating to $\sim$ 33\% of the targets being possibly active in this region. At this distance the sublimation would be dominated by volatiles such as CO or CO$_{2}$ \emph{if} sublimation was the driver behind the creation of the coma for these targets \citep{2017PASPWomack}. For an alternative to sublimation, other processes are also possible as discussed in \citet{2015P&SSNordheim} and \citet{2020JGREHergenrother}.

In Section~\ref{target_selection}, we discussed the parameter space used to select our targets, noting the presence of 143 catalogued active comets and 153 catalogued asteroids in this region. Based on our activity search in this study, if we multiply the active fraction by the total population, this would imply that at least $\sim 30$ catalogued asteroids may exhibit low-level activity or sublimation in this region. However, this inference only holds true if our small sample is representative of the broader object population in this region. 

From our sample of asteroids, it is perhaps most unexpected that we found potential activity on 2020 RX133 at $R_{h} = 4.56$ au, as this asteroid has one of the most distant perihelia at $q = 4.51$ au, one of the lowest eccentricities at $e = 0.11$, and the largest semi-major axis of all the asteroids in our sample of $a = 5.05$ au. While activity detection at such a distance is not uncommon, 2020 RX133 stood out as one of the least likely candidates for activity, given its orbital parameters. In fact, it was originally classified as one of our tertiary targets, indicating a lower priority for observation. Nevertheless, weakly active objects such as MBCs can show PSFs close to or the same as stars, for example MBC 133P/Elst–Pizarro \citep{1996IAUCElst, 2010MNRASHsieh} and MBC (248370) 2005 QN173 \citep{2021ApJHsieh}. Also see \citet{chandlerActiveAsteroidsCitizen2024} for further examples of active asteroids with stellar-like PSF's.

Based on the detection of potential low-level activity in these three target asteroids, we conclude that they are possible cometary nuclei undergoing weak sublimation. Should this activity be confirmed in future studies, for example via monitoring with the Vera C. Rubin Observatory \citep{2019ApJIvezic}, re-classification as comets would be appropriate. 

\subsection{Possible First Evidence of Activity in a Jupiter Trojan}\label{trojan}
\begin{figure}[h]
    \centering
    \epsscale{1.25}
    \plotone{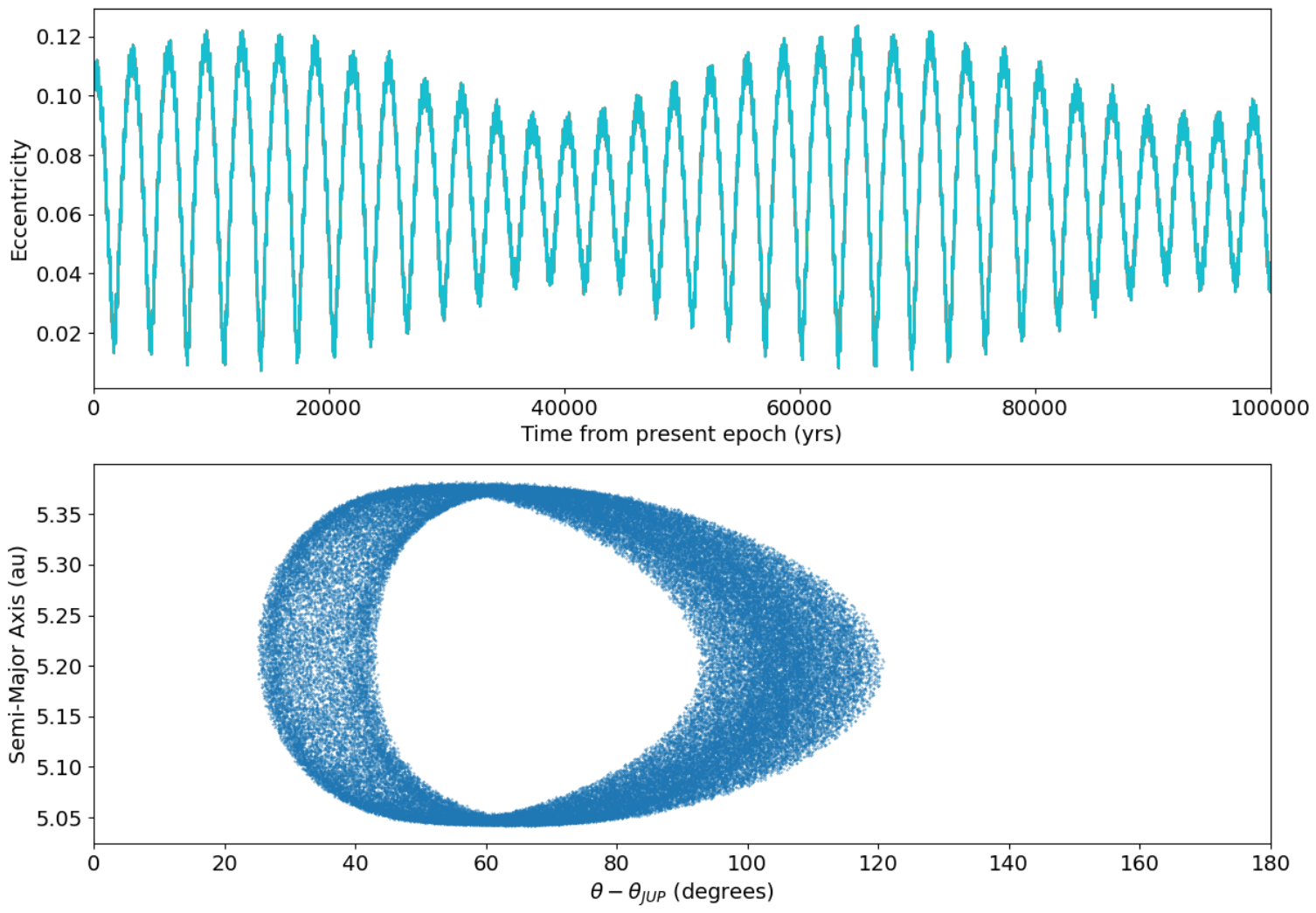}
    \caption{Dynamical evolution of 10 clones of 2020 RX133 over the next $10^5$ years. This is a subsample plotted for clarity from the full 100-clone study over $\pm10^6$ years described in the text. The full simulation showed the same behavior. Top: Evolution of eccentricity with time. Bottom: Phase plot of semi-major axis versus orbital longitude with respect to Jupiter. The small oscillations in eccentricity and the libration around the L4 point at $(\theta-\theta_{JUP})=60^\circ$ demonstrate this object is currently a stable Jupiter Trojan.}
    \label{fig:RX133_evolution}
\end{figure}

2020 RX133 is classed as a Jupiter Trojan in both the JPL Small-Body Database (SBD)
\footnote{\url{https://ssd.jpl.nasa.gov/tools/sbdb_lookup.html}}
and the ASTDyS-2
\footnote{\url{https://newton.spacedys.com/astdys2/}}
databases. However, the example of P/2019 LD2 shows weakly active  objects near Jupiter can have osculating elements similar to Trojan asteroids while undergoing strong orbital evolution 
\citep{2020RNAASKareta, 2021IcarHsieh}. To confirm its Trojan nature, we took the current orbital elements from the JPL SBD (epoch JD 2461000.5) and generated 100 clones by sampling the associated covariance matrix. We used REBOUND \citep{rebound} with the TRACE integrator \citep{reboundtrace} to numerically simulate their dynamical evolution due to the four major planets $\pm10^6$ years from the present epoch. The results are illustrated in Figure~\ref{fig:RX133_evolution}. All clones showed only small oscillations in eccentricity and semi-major axis, and all exhibited the classic libration in orbital longitude and semi-major axis exhibited by Jupiter Trojans \citep{1999ssdMurray}.

Jupiter Trojans orbit $\sim 5.2$ au from the Sun around the L4 and L5 Jupiter Lagrangian points and are thought to have formed in the outer solar system where they almost certainly accreted volatile ices in the process \citep{2023SSRvBottke}. Jupiter Trojans may possess
volatile-rich interiors with water-ice able to exist $\sim$ 10 m below the surface under a dust mantle \citep{2014IcarGuilbert-Lepoutre}. Recent James Webb Space Telescope observations also indicate that CO$_2$ may be a major sub-surface constituent \citep{2024PSJWong}. To date, remote observations have failed to detect definitive cometary coma or tails on Jupiter Trojans. They are the targets of NASA’s Lucy mission \citep{2021PSJLevison} which will soon provide in-situ searches for evidence of activity \citep{2025PSJStern}. In this context, our detection of candidate activity in 2020 RX133 represents a potentially significant first indication that Jupiter Trojans can display comet-like behavior. Further observations are required, however, to confirm the activity in 2020 RX133.

\subsection{Activity of the known comets}
Two of the known comets in our sample—286P/Christensen and 291P/NEAT—have previously published $Af\rho$ measurements. We compared their dust production rates to evaluate consistency across different observational methodologies. They were observed by the Asteroid Terrestrial-impact Last Alert System (ATLAS) and the $Af\rho$ for each was measured by \citet{2025PSJGillan}. In this work, we have calculated $Af\rho$ in terms of magnitudes, shown in Equation~\ref{eq:AfpMag}.

\begin{equation} \label{eq:AfpMag}
Af\rho = \frac{4 \, R_{h}^{2} \, \Delta^{2}}{\rho} \cdot 10^{0.4(m_{\odot} - m)}
\end{equation}

Here, $m_{\odot}$ and $m$ are the apparent magnitudes of the Sun and the comet respectively in the same filter. To inter-compare the two studies, we used the conventional $\rho = 10,000$ km aperture on the sky and used the photometric transformations given in \citet{2018PASPTonry} to convert the ATLAS broadband measurements to the Pan-STARRS photometric system where $r(PS1)= o(ATLAS)+0.15$, assuming a median color in the ATLAS filters of of $(c-o)=0.42$.

We calculated $Af\rho$ using two different magnitudes: the total magnitude from aperture photometry and the coma magnitude. Both were used to assess any differences arising from the choice of magnitude. 


\subsubsection{286P/Christensen}
We observed 286P/Christensen on 2022 November 17 at $R_{h} = 2.72$ au in the Sloan $r'$-band, using $28 \times 65$ second exposures. Calibrating to the \emph{PS1} photometric system and converting our magnitude measurements to $Af\rho$, we obtained $A(0^\circ)f\rho = 11 \pm 0.3$ cm using the aperture photometry magnitude, and $A(0^\circ)f\rho = 5 \pm 2$ cm using the coma magnitude.

\citet{2025PSJGillan} reported measurements from the ATLAS \emph{o}-band, which we converted to the $r(PS1)$ band. ATLAS recorded four observations at $R_{h} = 2.72$ au on 2022 November 19 at $R_{h} = 2.72$ au. These yielded $A(0^\circ)f\rho$ from $11 \pm 3$ cm to $20 \pm 4$ cm within a 10,000 km aperture. The large spread in ATLAS magnitudes -- and therefore in the resulting $Af\rho$ values -- is likely due to the comet's faint magnitude $\sim 19.5$, which is close to the ATLAS detection limit.

The results from this study (INT) and ATLAS agree within $1\sigma$ when using the total magnitude, and within $2\sigma$ when using the coma magnitude. In the ATLAS images, 286P appears very faint with no clear signs of activity,  due to the short 30 second exposures. In contrast, the longer exposure times and higher signal-to-noise ratio in the INT images reveal a clear coma surrounding the comet, showing obvious activity at the time of observation.


\subsubsection{291P/NEAT}
We observed 291P/NEAT on 2022 November 17 at $R_{h} = 2.82$ au with the INT. Using a combined image of $8 \times 120$ second exposures and calibrating to the \emph{PS1} system, we measured $A(0^\circ)f\rho = 21 \pm 1$ cm using the aperture photometry magnitude, and $A(0^\circ)f\rho = 17 \pm 2$ cm using the coma magnitude.

We compared these results with the measurements reported in \citet{2025PSJGillan} at the same heliocentric distance and date, converting the ATLAS \emph{o}-band magnitudes to the $r(PS1)$ band. The corresponding $A(0^\circ)f\rho$ values ranged from $22 \pm 5$ cm to $31 \pm 5$ cm within a 10,000 km aperture.

The two studies agree within $1\sigma$ when using both the total magnitude and the coma magnitude. As with 286P, the longer exposure times and higher signal-to-noise in the INT images reveal a clear coma surrounding 291P, which is not as apparent in the shorter ATLAS exposures. The agreement between datasets is encouraging but limited by the relatively small number of measurements at these faint magnitudes. Future higher-cadence or multi-band observations could help further constrain variability and dust properties for both 286P and 291P.

\section{Conclusions}\label{sec:conclusion}
We used the Wide Field Camera on the Isaac Newton Telescope at the Roque de los Muchachos Observatory on a three-night observing run from 2022 November 15 -- 2022 November 17. We observed known asteroids that were primarily on Jupiter-crossing orbits beyond the detection limit of typical survey imaging with semi-major axes in the range 4.05 $<$ a $<$ 5.05 au and $T_J$ $\leq$ 3.05, to search for low-level activity. We observed 16 asteroids and 7 known comets (both long-period and short-period) using the Sloan $g'$ and $r'$ filters.\\

\begin{enumerate}[nolistsep]

\item We compared the FWHM of the Moffat fit for the asteroids in median-combined images to the FWHM of the Moffat fit for the stars. We found that this method could determine clearly active objects from obvious inert objects, and may serve as a quick reference. It was not successful in identifying objects with potential very-low activity.

\item We measured 17 surface brightness profiles for 14 of our asteroidal targets. 2011 WM183, (669525) 2012 XO144, and 2020 RX133 showed surface brightness profiles that were consistent with low-level activity, equating to $\sim19$\% of our sample. When we considered the heliocentric distance range of the asteroids at the time they showed activity, this fraction increased to 33\% of the targets in the $3.16 \leq R_{h} \leq 4.56$ au region, and therefore it is possible to infer that at least $\sim30$ known asteroids with $T_J \leq 3.05$ and in the $4.05 < a < 5.05$ au parameter space may potentially exhibit low-level activity. Based on this, a cometary re-designation would be appropriate should these detections be confirmed. 

\item 2020 RX133 is a Jupiter Trojan asteroid, which makes the possible activity detection on this object particularly intriguing. If confirmed, this would be the first observation of activity in the Jupiter Trojan population.

\item For the three potentially active targets, we measured the coma contribution to our $r'$-band magnitudes. We also measured the relative dust production in terms of $A(0^\circ)f\rho$ for the active targets and known comets, where the cataloged comets typically showed larger dust production rates. We determined the nuclear magnitudes and radius estimates for the three potentially active asteroids of $r_{n} = 1.8 \pm 0.2$ km, $r_{n} \leq 0.8$ km, and $r_{n} \leq 0.5$ km for (669525) 2012 XO144, 2011 WM183, and 2020 RX133 respectively.

\item Finally, we measured the $(g-r)_{PS1}$ color indices for the majority of our asteroid targets. We found that a median  color index of $(g-r)_{PS1} = 0.52 \pm 0.13$ which indicated that most of our colors were similar with those expected for D-type asteroids within 1$\sigma$.
\end{enumerate}

Overall, this study demonstrated that although activity on objects evolving inward past Jupiter may remain undetected with typical wide-field survey imaging, faint dust comae can be detected through deep targeted imaging.

\begin{acknowledgments}
We thank the two anonymous reviewers for useful
comments and feedback, both of which have improved this
paper. Based on observations made with the Isaac Newton telescope operated on the island of La Palma by the Isaac Newton Group of Telescopes in the Spanish Observatorio del Roque de los Muchachos of the Instituto de Astrofísica de Canarias. 
Based on observations obtained with the Apache Point Observatory 3.5-meter telescope, which is owned and operated by the Astrophysical Research Consortium. Observations made use of Astrophysical Research Consortium Telescope Imaging Camera (ARCTIC) imager \citep{huehnerhoffAstrophysicalResearchConsortium2016a}. ARCTIC data reduction made use of the \texttt{acronym} software package \citep{l.weisenburgerAcronymAutomaticReduction2017}.

A.F.G. acknowledges support from the Department for the Economy (DfE) Northern Ireland postgraduate studentship scheme. A.F. acknowledges support from STFC award ST/T00021X/1. 

This research received support through Schmidt Sciences. 
C.{}O.{}C.\ acknowledges support from the DiRAC Institute in the Department of Astronomy at the University of Washington. The DiRAC Institute is supported through generous gifts from the Charles and Lisa Simonyi Fund for Arts and Sciences, Janet and Lloyd Frink, and the Washington Research Foundation.

World Coordinate System corrections were facilitated by \textit{Astrometry.net} \citep{langAstrometryNetBlindAstrometric2010}.
This research has made use of 
 NASA's Astrophysics Data System, 
 data and/or services provided by the International Astronomical Union's Minor Planet Center, \texttt{SAOImageDS9}, developed by Smithsonian Astrophysical Observatory \citep{joyeNewFeaturesSAOImage2006}. 
This study also made use of the National Aeronautics and Space Administration (NASA)/Jet Propulsion Laboratory (JPL) Small-Body Database (2025).
\end{acknowledgments}

\emph{Software:}
{\tt Python} (https://www.python.org),
{\tt{Astropy} \citep{{Astropy2013}, {price2018astropy}, {2022ApJAstropy}}, 
{\tt JPL Horizons} \citep{giorginiJPLsOnLineSolar1996},
{\tt Jupyter Notebook} \citep{Kluyver2016jupyter}, 
{\tt Matplotlib} \citep{Hunter:2007}, 
{\tt Numpy} \citep{harris2020array}, 
{\tt Pandas} \citep{mckinney2010data}, 
{\tt Photutils} \citep{larry_bradley_2022_6825092}, 
{\tt SkyBot} \citep{berthierSkyBoTNewVO2006},
{\tt{Rebound} \citep{{rebound}}.


\bibliography{main}{}
\bibliographystyle{aasjournal}

\end{document}